\documentclass[review]{elsarticle}

\usepackage{lineno,hyperref}
\modulolinenumbers[5]

\usepackage{graphicx}
\usepackage{dcolumn}
\usepackage{bm}
\usepackage{multirow}
\usepackage{textcomp}
\usepackage[]{subfigure}

\usepackage{booktabs}
\usepackage{array}
\usepackage{pgf}
\usepackage{graphicx}
\usepackage[absolute,overlay]{textpos}
\usepackage{calc}
\usepackage{amsmath}
\usepackage{amsfonts}
\usepackage{amssymb}
\usepackage{textcomp}
\usepackage{tikz}
\usepackage{gensymb}
\usepackage{color,soul}
\usepackage{xcolor}

\newcommand{\commentCWS}[1]%
{\textsf{\textcolor{blue}{#1$^{\mathrm{CWS}}$}}}
\usepackage[colorinlistoftodos]{todonotes} 

\begin{document}

\begin{frontmatter}

\title{Robust Property Design Under Composition Uncertainty}

\author[ubc]{C. W. Sinclair}
\ead{chad.sinclair@ubc.ca}

\author[ircp]{L. Lilensten}
\author[constellium]{F. Mas}
\author[mateis]{M. Perez}

\cortext[cor1]{Corresponding author}

\address[ubc]{
Department of Materials Engineering,
The University of British Columbia,
309-6350 Stores Road,
Vancouver, BC, Canada
}

\address[ircp]{
Chimie ParisTech, Universit\'e PSL, CNRS, Institut de Recherche de Chimie Paris, 75005 Paris, France
}

\address[constellium]{
Constellium C-TEC, Center of Technology,
Parc Economique Centr'alp,
CS10027, 38341 Voreppe Cedex, France
}

\address[mateis]{
Univ. Lyon, INSA Lyon, UCBL,
MATEIS, UMR CNRS 5510,
Villeurbanne, France
}

\begin{abstract}
Increasing the use of recycling in alloy production introduces the potential for greater variability in chemical composition, challenging conventional approaches to materials design and processing that optimize properties for a nominal chemistry. Here, we demonstrate an alternative approach in which \emph{robustness} to composition variability, expressed through the resulting spread in mechanical properties, is treated as a design objective. Using a precipitation-strengthening model coupled to a computationally efficient statistical surrogate, we investigate how heat treatment influences the sensitivity of yield strength to fluctuations in Mg and Si content in an Al–Mg–Si alloy.  We demonstrate that processing conditions maximizing mean strength do not necessarily minimize property variability, and that alternative aging treatments can substantially improve robustness with relatively
small penalties in strength. A multi-objective optimization framework identifies the trade-offs between these competing objectives. These results suggest that designing for robustness, rather than optimizing nominal performance alone, offers a complementary strategy to tighter composition control for accommodating heterogeneous recycled feedstocks.
\end{abstract}


\end{frontmatter}

\date{\today}

\section{Introduction}

Robustness, the ability of a system to maintain performance in the presence of perturbations and uncertainty, is a defining characteristic of many biological systems \cite{Stelling2004RobustnessFunctions,Kitano2004BiologicalRobustness, Kitano2007TowardsRobustness,Hamant2024Robustness}. Such systems rarely maximize a single measure of performance under one nominal condition. Instead, heterogeneity, redundancy and other apparently suboptimal features can increase resilience to changing conditions. More generally, Grumbach and Hamant have argued that deliberately sacrificing local or short-term optimality can provide a route to greater system-level resilience
\cite{Grumbach2020HowSystems}. This principle has a direct analogue in robust engineering design, where the objective is not necessarily to maximize nominal performance, but to maintain acceptable performance in the presence of unavoidable uncertainty \cite{Beyer2007RobustSurvey}. Climate change has brought this challenge into sharper focus in the design of buildings and infrastructure, where uncertain future operating conditions increasingly demand strategies that remain effective across a range of possible scenarios rather than being optimized for a single expected outcome \cite{Hallegatte2009StrategiesChange}.

In materials engineering, growing recognition of the environmental impacts of materials production and consumption is similarly motivating a shift toward circular strategies that prioritize material recovery and reuse. Aluminium alloys provide a particularly important example, given the substantial energy requirements of primary aluminium production and the potential benefits of recycling \cite{Cullen2013MappingGoods,Raabe2021SustainableAlloys,Desing2026ClimatePerspectives}. An emerging research challenge is therefore how to design alloys and processing routes that can accommodate the compositional heterogeneity of mixed recycled feedstocks \cite{Raabe2021SustainableAlloys,Gaustad2010DesignModification,Krall2026DirectVehicles}. This represents a departure from conventional alloy development, where improvements in primary-metal production, alloying and process control have enabled increasingly precise control over composition and processing conditions. Computational alloy and process design have consequently focused predominantly on optimizing properties for a nominal composition. Mixed post-consumer aluminium scrap, however, is intrinsically heterogeneous. Recent characterization of shredded end-of-life automotive scrap (``twitch''), for example, reveals mixtures of multiple wrought alloy families and substantial batch-to-batch variations in Si, Cu, Mg, Zn and other alloying elements \cite{Shankar2025UpcyclingTwitch,Krall2026DirectVehicles}. Conventional recycling therefore relies heavily on sorting, dilution and composition adjustment to recover materials suitable for demanding wrought-alloy applications.

An alternative to tighter control of feedstock chemistry is to ask whether alloys and processing routes can be designed to accommodate, rather than eliminate, some of
this compositional variability. This question is particularly important for precipitation-hardened alloys because composition affects not only the magnitude of strengthening but also precipitation kinetics. A distribution of starting compositions subjected to a common heat treatment therefore generates a time-dependent distribution of mechanical properties. Consequently, the heat treatment that maximizes strength for a nominal composition need not maximize the mean strength of a compositionally heterogeneous population, minimize property variability, or provide the greatest probability of satisfying engineering requirements. In the present work, we consider a processing condition to be more robust when it produces a narrower distribution of yield strengths for a given distribution of starting compositions. Processing conditions that sacrifice some mean strength may therefore offer a more robust mechanical response.

Here, we investigate robust property design using an age hardenable Al–Mg–Si alloy as a model system. Although mixed recycled aluminum feedstocks may contain variations in numerous alloying and impurity elements, we focus on fluctuations in Mg and Si to establish the principles of this approach within a well-defined composition space. Composition fluctuations in Mg can typically be adjusted during aluminum processing, making it less susceptible to uncontrolled composition variations than many other elements. We have, however, included it alongside Si to examine how uncertainty in different solute concentrations influence precipitation strengthening and the robustness of mechanical properties. We develop and calibrate a physics-based precipitation-strengthening model for an alloy with a nominal composition of Al–0.69 wt.\% Mg–1.01 wt.\% Si, and couple it with a Gaussian-process surrogate to efficiently predict the evolution of yield-strength distributions under different aging conditions. We then use Pareto analysis and a weighted strength-variability objective to identify heat treatments that balance mean strength against property consistency. In doing so, we demonstrate how robustness can be incorporated explicitly into materials property design, allowing compositional variability to be accommodated through processing rather than relying exclusively on tighter control of feedstock chemistry.

Precipitation kinetics were simulated using the class-based Kampmann--Wagner numerical (KWN) precipitation model PreciSo \cite{Perez2008ImplementationPrecipitation}, approximating the strengthening $\beta''$ phase in Al--Mg--Si alloys using Mg$_5$Si$_6$ stoichiometry. The model predicts the evolution of matrix solute concentrations, precipitate size distributions and precipitate volume fraction during aging. These quantities were coupled to a strengthening model incorporating solid-solution and precipitation hardening contributions. Solid-solution strengthening by Mg and Si was described using a Labusch-type relationship, with coefficients taken from \cite{Leyson2012SoluteAlloys}, while precipitation strengthening was calculated using the approach of \cite{deVaucorbeil2013TheAlloys}. The interfacial energy ($\gamma$), critical radius ($r_c$) and baseline strength ($\sigma_0$) were determined by jointly fitting the predicted yield-strength evolution to experimental aging curves at 160, 180 and 200$^\circ$C reported by Bao \cite{Bao2021Ageing}. A global optimization procedure employing a robust Cauchy loss function was used to obtain a single parameter set describing all three aging temperatures. The optimized parameters are summarized in Table~\ref{tab:model_parameters}, with details of the optimization procedure and comparison with experimental measurements provided in the Supplementary Material.

\begin{table}[ht!]
\centering
\caption{Nominal alloy composition, composition ranges used for Gaussian-process (GP) training, aging temperatures, and calibrated precipitation-strengthening model parameters. Here $r_c$, $\gamma$ and $\sigma_0$ are fitting parameters.}
\label{tab:model_parameters}
\begin{tabular}{ll}
\toprule
Parameter & Value \\
\midrule
Nominal composition
    & Al--0.69 wt.\% Mg--1.01 wt.\% Si \\
GP training range, Mg
    & 0.50--0.95 wt.\% \\
GP training range, Si
    & 0.65--1.25 wt.\% \\
Aging temperatures
    & 160, 180, 200 $^\circ$C \\
Interfacial energy, $\gamma$
    & 0.121 J\,m$^{-2}$ \\
Critical radius, $r_c$
    & 4.21 nm \\
Baseline strength, $\sigma_0$
    & 19.9 MPa \\
\bottomrule
\end{tabular}
\end{table}

Direct evaluation of the coupled precipitation–strengthening model over the large Monte Carlo ensembles required for robustness analysis is computationally expensive. To make these calculations computationally tractable, we constructed a statistical surrogate model that approximates the relationship between alloy composition, aging conditions and predicted yield strength. Specifically, we employed Gaussian-process (GP) regression, a nonlinear regression method that interpolates between training calculations while also providing an estimate of the uncertainty associated with its predictions. Separate GP surrogates were constructed at each aging temperature to predict yield strength as a function of Mg content, Si content and aging time. The initial training sets comprised 25 alloy compositions selected using a space-filling design spanning the Mg–Si composition ranges given in Table~\ref{tab:model_parameters}. These were subsequently refined through 20 active-learning iterations, using the GP predictive uncertainty to identify compositions for which additional precipitation–strengthening calculations would be most informative. Each sampled composition contributed a complete aging curve to the training data. Following validation against independent precipitation–strengthening calculations, the GP surrogates were used for the subsequent uncertainty-propagation analyses. 

The computational efficiency arising from the use of the GP surrogates can be seen from the $1.35\times10^6$ yield-strength evaluations used for uncertainty-propagation calculations presented below. These were completed in approximately 76~s using the trained surrogate. Based on measured execution times of 3.8–4.5~s per complete PreciSo aging curve, the equivalent direct calculations would require an estimated $1.6\times10^3$ CPU hours, corresponding to a speedup of between 4 and 5 orders of magnitude, excluding the one-time cost of constructing the surrogates. Further details of the surrogate construction and validation are provided in the Supplementary Material.

To examine the influence of composition variability, Mg and Si contents were independently sampled from Gaussian distributions centred on the nominal composition (Table~\ref{tab:model_parameters}). Both elements were assigned the same standard deviation, $\mathcal{S}$, which was varied between 0.01 and 0.06 wt.\%. The imposed standard deviation characterizes the magnitude of the composition fluctuations and is distinct from the composition ranges used to train the GP surrogate models. For comparison, the commercial AA6082 alloy specification permits Mg and Si contents of 0.6–1.2 wt.\% and 0.7–1.3 wt.\%, respectively though the composition range of an individual industrial producer will be currently much narrower than this. At the largest imposed variability, approximately 93\% of compositions drawn from the untruncated Gaussian distributions satisfy both of these industrial specification limits. The calculations therefore explore composition fluctuations comparable to those accommodated within existing commercial alloy specifications. For each aging temperature and time, the GP surrogate was evaluated over the sampled compositions to obtain the corresponding distribution of yield strength. These distributions were characterized by their mean, $\overline{\sigma}_y$, and their 90\% width,

\begin{equation}
W_{90} = q_{95}(\sigma_y)-q_{05}(\sigma_y)
\end{equation}

where $q_{05}$ and $q_{95}$ denote the fifth and ninety-fifth percentiles, respectively. The width $W_{90}$ provides a measure of composition-induced strength variability without assuming that the resulting strength distributions are Gaussian. For a fixed level of composition scatter, a smaller $W_{90}$ therefore indicates
greater robustness of the mechanical response to composition fluctuations.

\begin{figure}[ht!]
    \centering
    \includegraphics[width=\columnwidth]{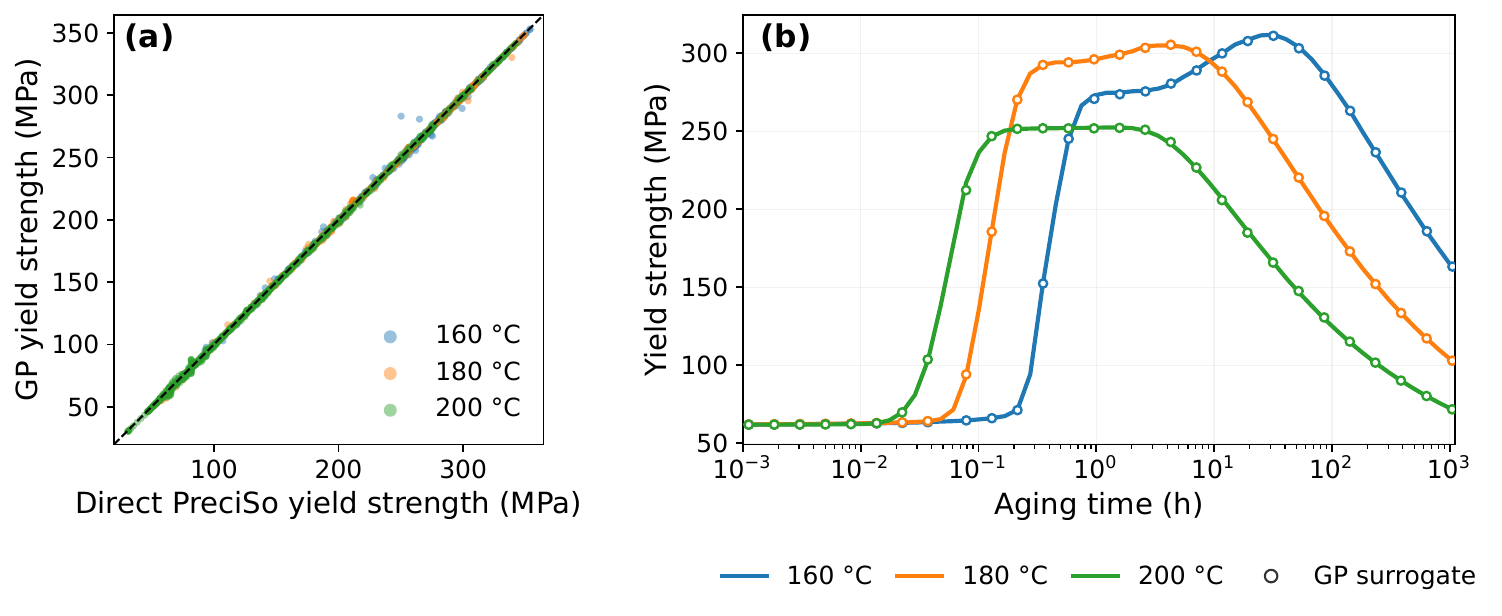 }
    \caption{Validation of the Gaussian-process (GP) surrogates against direct PreciSo precipitation--strengthening calculations at 160, 180 and 200$^\circ$C. (a) Parity plot comparing GP-predicted and directly calculated yield strengths for compositions excluded from surrogate training. (b) Comparison of the predicted aging responses for an independent validation composition of Al--0.7406 wt.\% Mg--1.0158 wt.\% Si. Solid lines represent direct precipitation--strengthening calculations, while symbols indicate GP predictions.}
    \label{fig1}
\end{figure}

The accuracy of the GP surrogates is assessed in Fig.~\ref{fig1} using compositions excluded from the
training process. Figure~\ref{fig1}(a) compares the yield strengths predicted by the GP surrogates with those obtained directly from the coupled precipitation--strengthening model at 160, 180 and 200$^\circ$C. Across 1920 validation points at each temperature, the mean absolute errors are below 0.32 MPa, demonstrating excellent agreement over the full range of predicted strengths. The ability of the surrogates to reproduce the time-dependent aging response is further illustrated in Fig.~\ref{fig1}(b),
where predictions for an independently selected composition closely follow the direct model calculations through underaging, peak aging and overaging. 

\begin{figure}[ht!]
    \centering
    \includegraphics[width=\columnwidth]{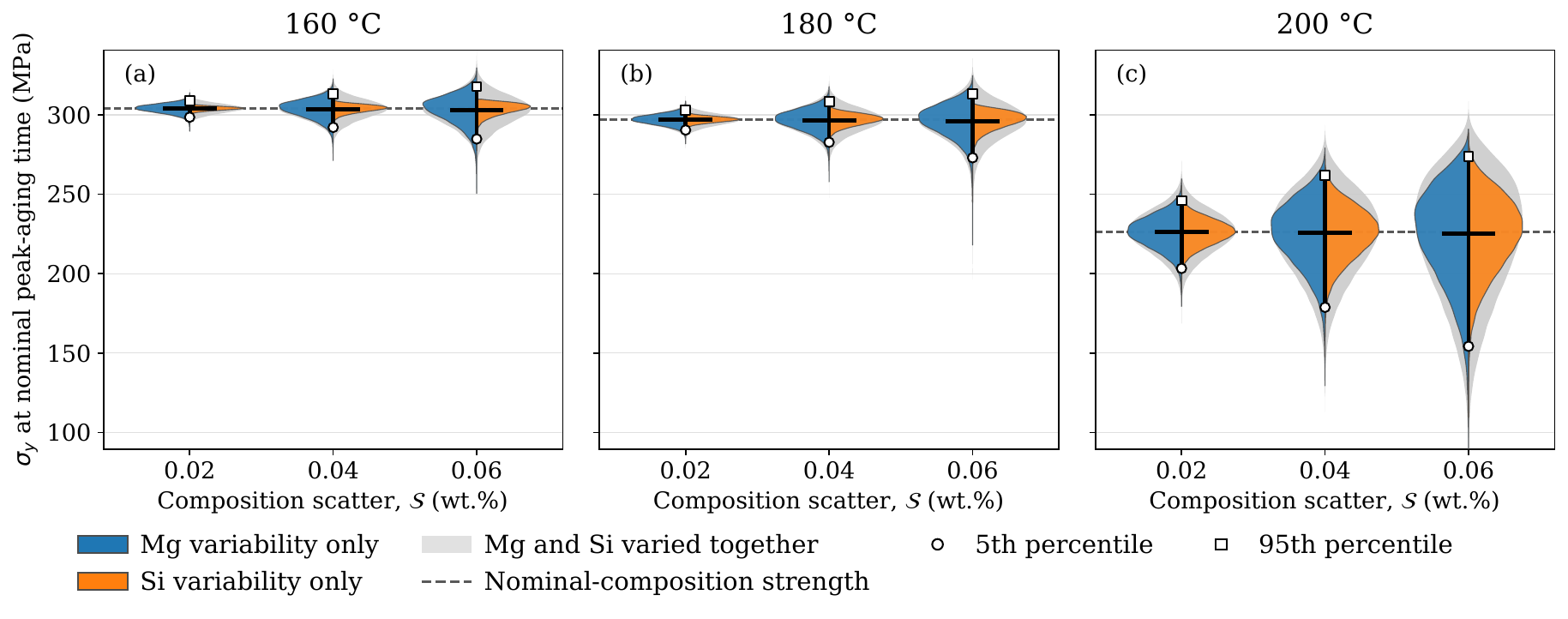}
    \caption{Influence of composition variability on the distribution of yield strength ($\sigma_y$) at the nominal-composition peak-aging time for (a) 160, (b) 180 and (c) 200$^\circ$C. Split violin plots show the distributions obtained when only Mg (blue) or only Si (orange) is varied, while the grey distributions represent simultaneous independent variations in both elements. The imposed composition scatter $\mathcal{S}$, is the standard deviation of each varied element in wt.\%. Circles and squares
    indicate the 5th and 95th percentiles, respectively, of the combined Mg--Si distributions. Horizontal dashed lines indicate the yield strength predicted for the nominal composition.
    }
    \label{fig2}
\end{figure}

Using the validated GP surrogates, Monte Carlo uncertainty propagation was performed to quantify how variations in Mg and Si content influence the distribution of yield strength. Rather than predicting a single strength for a nominal alloy composition, this approach provides a distribution of strengths associated with a prescribed level of composition variability. Figure~\ref{fig2} presents the resulting yield-strength distributions for aging treatments corresponding to the peak strength of the nominal alloy composition (Al–0.69 wt.\% Mg–1.01 wt.\% Si) at each of the three temperatures. For each temperature, the aging time is held fixed at this reference value while the Mg and Si concentrations are varied. Consequently, the sampled compositions are not necessarily evaluated at their individual peak-strength conditions. The concentrations were varied independently, with a common standard deviation, $\mathcal{S}$, ranging from 0.02 to 0.06 wt.\%. The individual contributions of Mg and Si variability are distinguished from the distributions obtained when both elements vary simultaneously.

Increasing composition variability produces progressively broader strength distributions, while their mean values remain relatively close to the strength predicted for the nominal composition. The magnitude of this broadening depends strongly on aging temperature. At 160$^\circ$C, the strength distribution remains comparatively narrow even at the largest composition scatter considered. The distributions become broader at 180$^\circ$C and substantially broader at 200$^\circ$C, where variations in Mg content contribute particularly strongly to the overall strength variability.

The distributions also become increasingly asymmetric as composition variability increases, developing a more pronounced tail toward lower strengths. This behaviour arises because the relationship between alloy composition and yield strength is nonlinear meaning that Gaussian fluctuations in Mg and Si content need not produce Gaussian strength distributions. Nevertheless, the central 90\% widths remain within approximately 2.5\% of those expected for Gaussian distributions having the same standard deviation. Thus, while the overall width of the strength distribution is well characterized by its variance, this measure alone does not capture the increasing
asymmetry toward lower strengths.

\begin{figure}[t]
    \centering
    \includegraphics[width=\columnwidth]{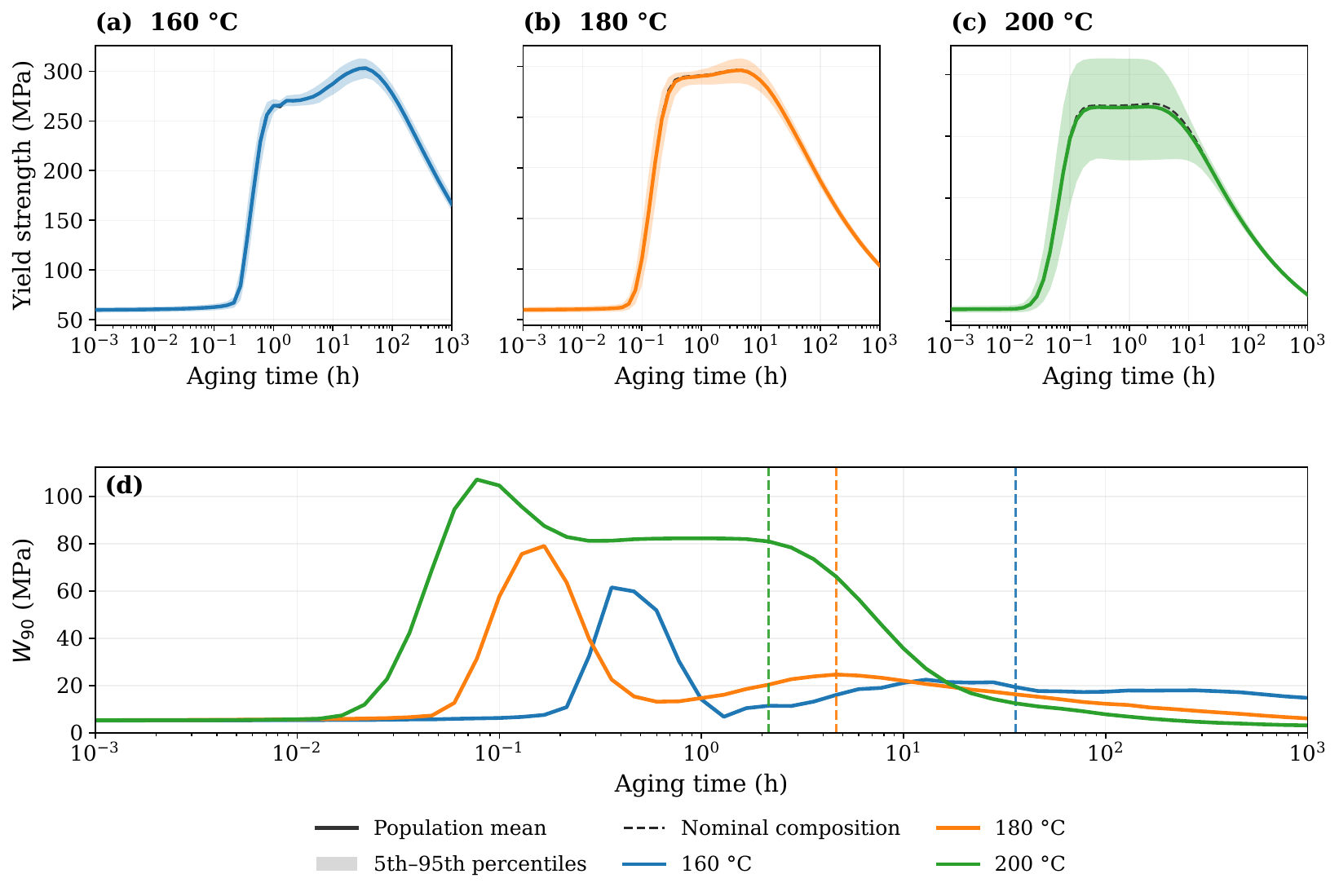}
    \caption{ Evolution of composition-induced yield-strength variability during isothermal aging at (a) 160, (b) 180 and
    (c) 200$^\circ$C. Independent fluctuations in Mg and Si content are imposed with standard deviations $\mathcal{S}_{\rm Mg}=\mathcal{S}_{\rm Si}=0.04$ wt.\%. Solid curves show the population mean yield strength, while shaded regions indicate the 5th--95th percentile interval. Black dashed curves represent the response of the nominal composition. (d) Corresponding evolution of the 90\% strength-distribution width, $W_{90}=q_{95}-q_{05}$.  Coloured vertical dashed lines indicate the nominal-composition peak-aging times at each temperature.}
    \label{fig3}
\end{figure}

Figure~\ref{fig2} illustrates the influence of composition variability at the nominal-composition peak-aging condition. However, the precipitation kinetics, and therefore the sensitivity of strength to composition, evolve continuously during aging. Figure~\ref{fig3}(a--c) shows the evolution of mean yield strength and the corresponding 5th--95th percentile intervals at 160, 180 and 200$^\circ$C, for a composition scatter of $\mathcal{S}=0.04$ wt.\%. The widths of these intervals change considerably during aging, even over periods when the mean strength varies relatively little.

This behaviour is quantified in Fig.~\ref{fig3}(d), which shows the evolution of the 90\% strength-distribution width, $W_{90}=q_{95}-q_{05}$. At each temperature, the onset of precipitation strengthening is accompanied by a pronounced increase in $W_{90}$, reflecting the sensitivity of the developing precipitation kinetics to differences in alloy composition. The subsequent evolution is strongly temperature dependent. At 160 and 180$^\circ$C, relatively sharp maxima in $W_{90}$ occur during the early stages of strengthening, followed by substantial reductions in variability. At 200$^\circ$C, the increase is considerably larger, and the strength distribution remains broad over much of the peak-strength regime before narrowing during overaging. The magnitude of this variability is particularly striking at 200$^\circ$C, where the width of the strength distribution becomes a substantial fraction of the mean yield strength. Thus, even relatively modest fluctuations in starting composition can produce large differences in mechanical response between alloys subjected to the same heat treatment. Importantly, the times corresponding to maximum strength and maximum variability do not generally coincide.

The greater strength variability at 200$^\circ$C can be understood in terms of the sensitivity of precipitation nucleation to composition-dependent supersaturation. As the aging temperature increases from 160 to 200$^\circ$C, the critical nucleus radius increases from approximately 0.37 to 0.50~nm, corresponding to an increase in the nucleation barrier from approximately 12 to 19.5~$k_BT$. At 200$^\circ$C, variations of $\pm 0.04$ wt.\% in Mg or Si produce changes of approximately 25--30\% in precipitate number density at the nominal peak-aging time, compared with much smaller changes at 160$^\circ$C. Despite these differences, the precipitate volume fraction varies by only a few percent at 200$^\circ$C. The resulting differences in precipitate number density therefore translate primarily into changes in mean precipitate radius and interparticle spacing, producing variations in precipitation strengthening of approximately $\pm 20$~MPa. This mechanism also helps explain the asymmetric strength distributions observed in Fig.~\ref{fig2}. Reductions in solute content increase the nucleation barrier and suppress the formation of precipitates, producing a coarser precipitate population and disproportionately large reductions in precipitation strengthening. The corresponding increases in strength for solute-rich compositions are smaller, giving rise to the extended low-strength tails.

\begin{figure}[th!]
    \centering
    \includegraphics[width=\columnwidth]{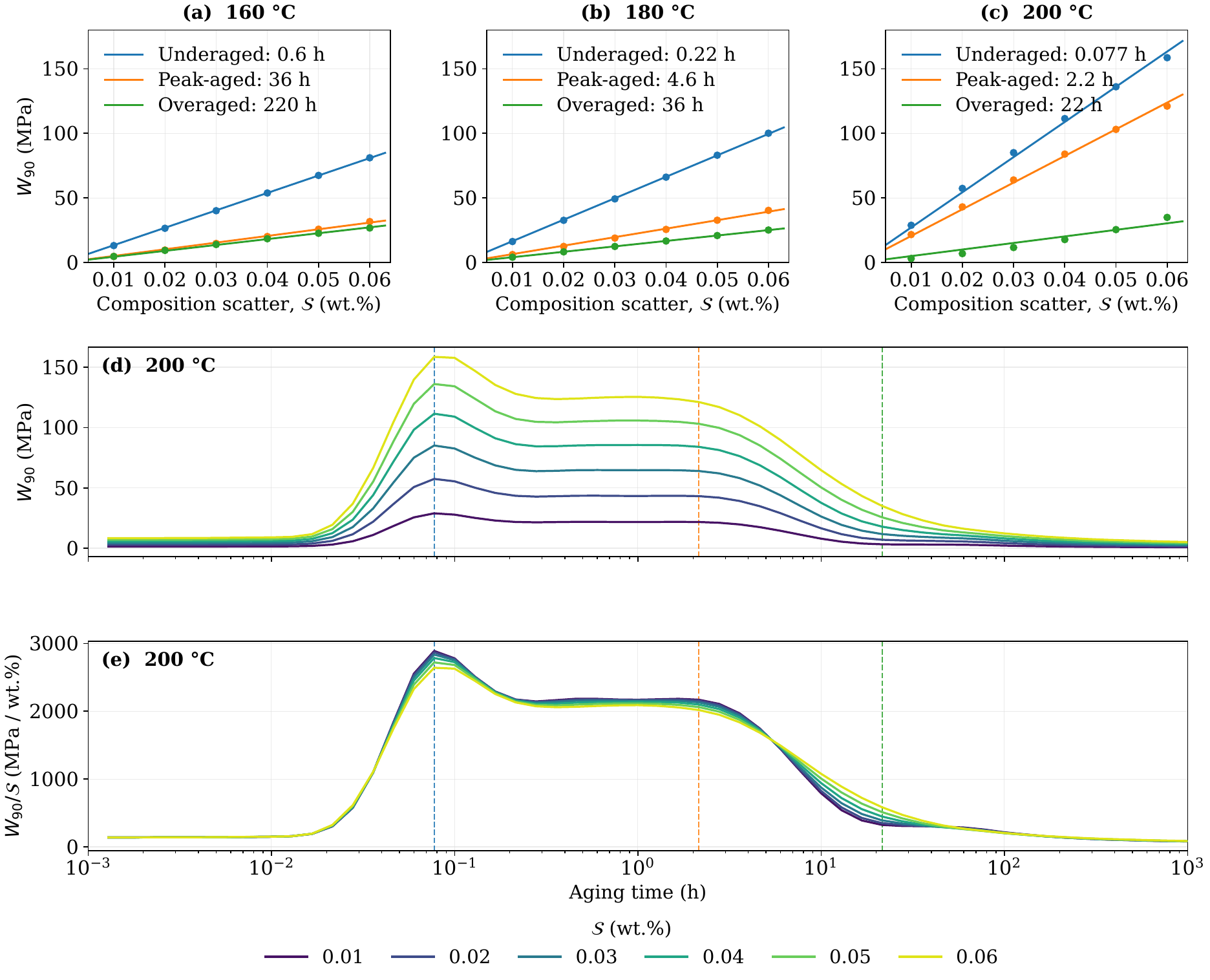}
    \caption{Scaling of composition induced yield strength variability with the imposed composition standard deviation, $\mathcal{S}$. (a–c) Dependence of the 90\% strength-distribution width, $W_{90}=q_{95}-q_{05}$, on $\mathcal{S}$ for underaged, peak-aged and overaged treatments at 160, 180 and 200$^\circ$C. The underaged and overaged conditions correspond to approximately 80\% of the peak strength of the nominal alloy composition at each temperature, with aging times held fixed during composition sampling. Symbols indicate Monte Carlo results and solid lines indicate linear fits. (d) Evolution of $W_{90}$ during aging at 200$^\circ$C for different values of $\mathcal{S}$. (e) Corresponding evolution of $W_{90}/\mathcal{S}$, illustrating the approximate linear scaling of strength variability with composition scatter. Vertical dashed lines indicate the underaged, peak-aged and overaged conditions examined in (c).}
    \label{fig4}
\end{figure}

An important further question is how the strength variability scales with the magnitude of the imposed composition fluctuations. Figure~\ref{fig4}(a–c) shows the dependence of $W_{90}$ on $\mathcal{S}$ for representative underaged, peak-aged and overaged treatments at each temperature. These conditions were selected based on the aging response of the nominal alloy composition, with the underaged and overaged treatments chosen to produce strengths close to 80\% of the corresponding peak strength. The aging time associated with each treatment is then held fixed while the compositions are varied, so that individual sampled alloys may occupy different stages of their precipitation response. Despite the nonlinear dependence of precipitation strengthening on composition, $W_{90}$ increases approximately linearly with $\mathcal{S}$ over the range investigated. The slopes, however, vary substantially with aging condition, with the underaged treatments exhibiting particularly strong sensitivity to composition fluctuations.

The approximate scaling is further illustrated in Fig.~\ref{fig4}(d), which shows the evolution of $W_{90}$ at 200$^\circ$C for different values of $\mathcal{S}$. Although the magnitude of the strength variability increases substantially with composition scatter, its temporal evolution remains qualitatively similar. Normalizing by $\mathcal{S}$ produces an approximate collapse of the curves (Fig.~\ref{fig4}(e)), suggesting that the variability can be expressed, to first order, as $W_{90}(T,t,\mathcal{S})\approx K(T,t)\mathcal{S}$, where $K(T,t)$ characterizes the sensitivity of strength variability to composition fluctuations.

This behaviour can be understood through a first-order expansion of yield strength about the nominal composition. For small composition fluctuations, the change in yield strength is approximated by the sum of the Mg and Si concentration fluctuations multiplied by their respective partial derivatives of yield strength. Since the two concentration fluctuations are independent Gaussian variables, each with standard deviation $\mathcal{S}$, the variance of yield strength is approximately $\mathcal{S}^2$ multiplied by the sum of the squared partial derivatives. Under this linear approximation, the resulting strength distribution is also Gaussian, with its 5th and 95th percentiles separated by 3.29 standard deviations. Consequently,

\begin{equation}
W_{90}\approx 3.29\,\mathcal{S} \left[ \left(\frac{\partial\sigma_y}{\partial c_{\rm Mg}}\right)^2+ \left(\frac{\partial\sigma_y}{\partial c_{\rm Si}}\right)^2 \right]^{1/2}
\label{eq:strength_variability_scaling}
\end{equation}

The numerical factor 3.29 corresponds to the separation between the 5th and 95th percentiles of a Gaussian distribution, expressed in units of its standard deviation. The derivatives are evaluated at the nominal composition and the specified aging condition. Thus, $K(T,t)$ is proportional to the magnitude of the composition gradient of yield strength, which evolves with the precipitate population during aging. Higher-order composition dependence can introduce asymmetry into the strength distributions, as observed in Fig.~\ref{fig2}, without substantially altering the leading-order linear scaling of $W_{90}$ with $\mathcal{S}$.

The preceding results demonstrate that the sensitivity of yield strength to composition variability can be modified through the selection of aging temperature and time. However, reducing this sensitivity may require sacrificing mean strength. To examine this trade-off, we consider the competing objectives of maximizing the population mean yield strength, $\overline{\sigma}_y$, and maximizing robustness by minimizing the width of the resulting strength distribution, $W_{90}$. A treatment is considered Pareto efficient if no other feasible treatment at the same temperature provides both a higher mean strength and a smaller $W_{90}$.

\begin{figure}[t]
    \centering
    \includegraphics[width=\columnwidth]{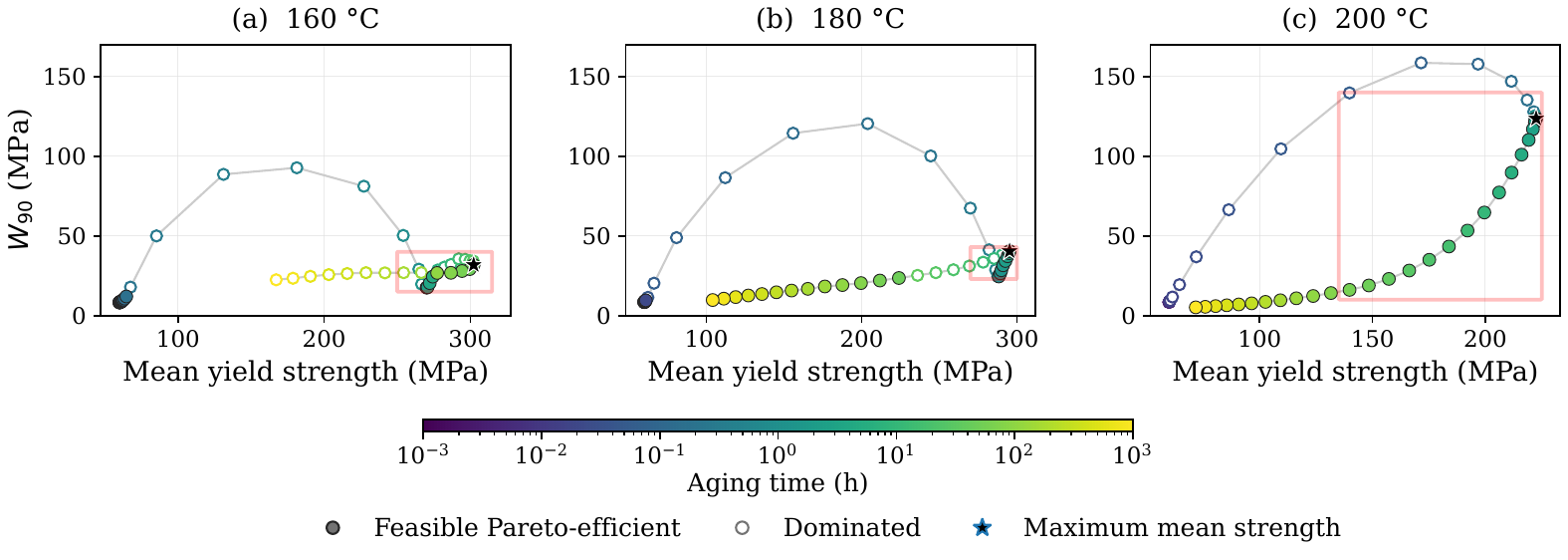}
    \caption{Trade-off between population mean yield strength, $\overline{\sigma}_y$, and composition-induced strength variability, $W_{90}$, during isothermal aging at (a) 160, (b) 180 and (c) 200$^\circ$C. Independent Mg and Si fluctuations are imposed with $\mathcal{S}=0.06$ wt.\%. Each point represents an aging treatment, with colour indicating aging time. Filled symbols identify temperature-specific Pareto-efficient treatments, while open symbols indicate dominated treatments. Stars mark the maximum mean strength at each temperature. The faint red boxes indicate the zoomed in regions shown in Figure \ref{fig6}.}
    \label{fig5}
\end{figure}

Figure~\ref{fig5} shows the resulting strength--variability trajectories at 160, 180 and 200$^\circ$C for independent Mg and Si fluctuations with $\mathcal{S}=0.06$ wt.\%. Each point represents a different aging time, indicated by its colour. 

The three temperatures exhibit markedly different strength/variability trade-offs. At 160 and 180$^\circ$C, the Pareto-efficient treatments close to the maximum mean strength occupy relatively narrow regions of the strength–variability space, with moderate underaging offering appreciable reductions in variability for relatively small sacrifices in strength. In contrast, at 200$^\circ$C, the Pareto frontier extends over a broader range during overaging, along which reductions in $W_{90}$ are accompanied by progressive losses in mean strength.

These differences demonstrate that there is no single isothermal aging strategy that simultaneously maximizes strength and minimizes its sensitivity to composition variability. Instead, the preferred treatment depends on the relative importance assigned to strength and property consistency. To illustrate this, we introduce a weighted objective function to be maximized \cite{Ashby2000Multi-objectiveSelection},

\begin{equation}
\mathcal{O} = \overline{\sigma}_y-\beta W_{90},
\label{eq:robust_objective}
\end{equation}

Here, $\beta$ is a dimensionless penalty assigned to the width of the strength distribution. At $\beta=0$, the objective reduces to maximizing mean yield strength, whereas increasing $\beta$ progressively favours treatments producing narrower strength distributions. The parameter $\beta$ therefore represents the relative importance assigned to reducing strength variability compared with maintaining mean strength. For example, $\beta=1$ implies that a reduction of 1 MPa in $W_{90}$ compensates for a loss of 1 MPa in mean yield strength.

For a fixed value of $\beta>0$, contours of the objective function (Eq.~\ref{eq:robust_objective}) are straight lines with slope $1/\beta$ in the mean strength–$W_{90}$ plane. Increasing the value of the objective function shifts these contours towards the lower right, corresponding to combinations of higher mean strength and lower variability. The optimal aging condition for a selected value of $\beta$ is therefore identified by the highest-value objective contour that intersects the feasible Pareto frontier. For a smooth, convex trade-off boundary, this corresponds to the point at which the objective contour is tangent to the frontier. More generally, for the discrete aging conditions considered here, the optimum is the Pareto-efficient condition that maximizes the weighted objective function.

\begin{figure}[t]
    \centering
    \includegraphics[width=\columnwidth]{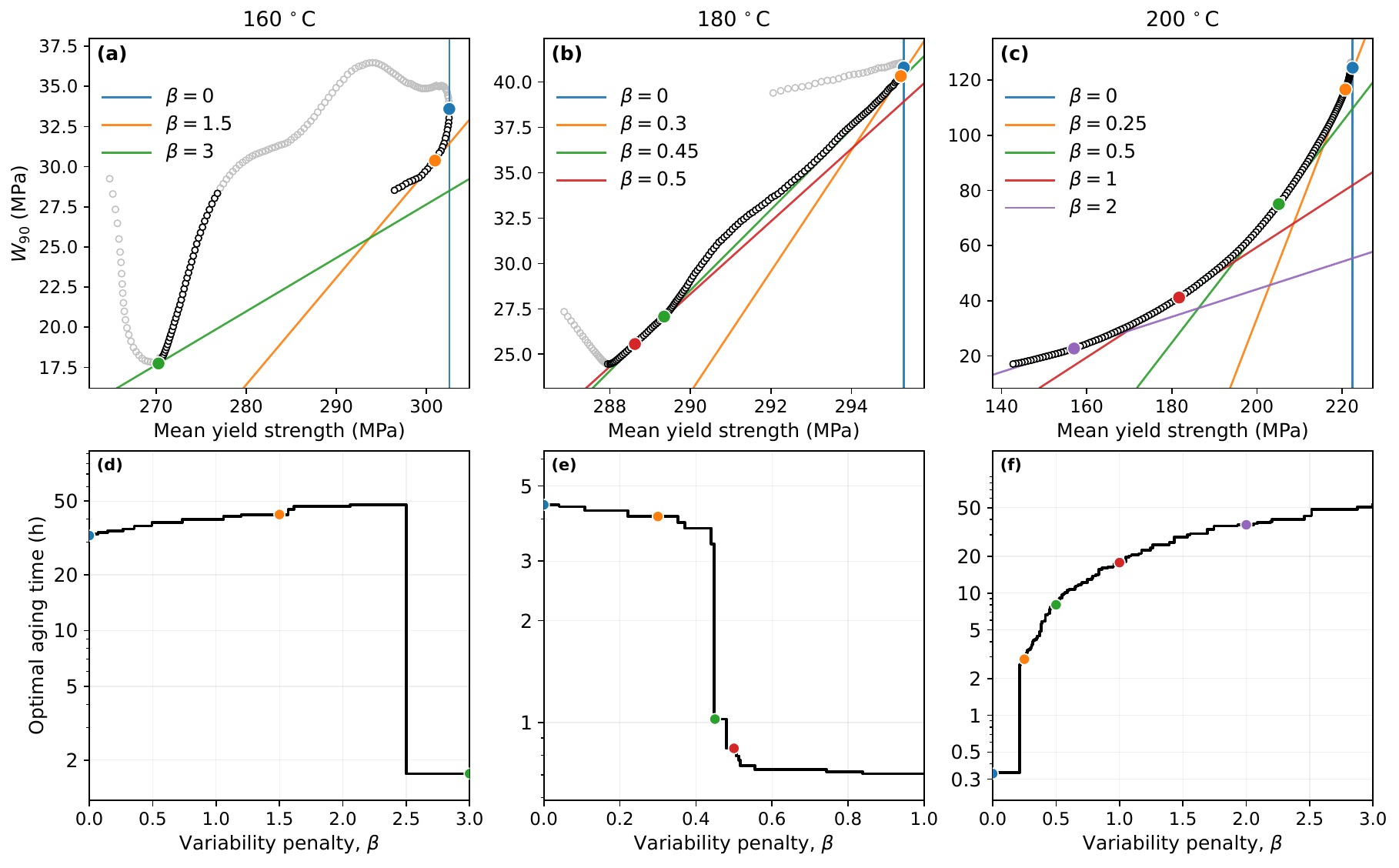}
    \caption{Selection of aging treatments balancing mean yield strength and composition-induced variability for independent Mg and Si fluctuations with $\mathcal{S}=0.06$ wt.\%. (a--c) Pareto-efficient strength--variability combinations at 160, 180 and 200$^\circ$C, respectively. Coloured solid lines indicate constant values of the weighted objective $\mathcal{O}=\overline{\sigma}_y-\beta W_{90}$ for selected variability penalties $\beta$; coloured symbols identify the corresponding optimal treatments. Open grey symbols indicate dominated or infeasible treatments. (d--f) Corresponding optimal aging time as a function of $\beta$ at each temperature.}
    \label{fig6}
\end{figure}

Figure~\ref{fig6} illustrates this optimization for the three aging temperatures at $\mathcal{S}=0.06$ wt.\%. The upper panels show the Pareto-efficient treatments together with contours of constant objective for selected values of $\beta$, while the lower panels show how the corresponding optimal aging times (tangents to paerto curve) vary with the variability penalty $\beta$.

At 160$^\circ$C, increasing $\beta$ initially favours slightly longer aging times, corresponding to modest overaging. This reduces $W_{90}$ with relatively little loss of mean strength. However, at $\beta\approx2.5$, the preferred treatment switches abruptly from a slightly overaged condition to a substantially underaged condition. Although this reduces $W_{90}$ to approximately 18~MPa, the mean strength falls to approximately 270~MPa. At 180$^\circ$C, increasing $\beta$ instead favours progressively shorter aging times, with a sequence of transitions toward more strongly underaged conditions near $\beta\approx0.45$--0.5. These changes reflect the shape of the Pareto frontiers: when distinct aging conditions offer competing strength--variability compromises, relatively small changes in the penalty assigned to variability can produce abrupt changes in the optimal treatment.

At 200$^\circ$C, the response is qualitatively different. Increasing $\beta$ produces a progressive increase in the optimal aging time, corresponding to increasingly overaged conditions. For example, increasing $\beta$ from zero to approximately two reduces $W_{90}$ from about 125 to 22~MPa, but also decreases the mean strength from approximately 220 to 160~MPa. Thus, while overaging provides a means of progressively reducing composition-induced variability at 200$^\circ$C, the accompanying loss of mean strength limits its attractiveness for applications requiring high yield strength.

Although Fig.~\ref{fig6} presents results for $\mathcal{S}=0.06$ wt.\%, the approximate linear scaling of $W_{90}$ with $\mathcal{S}$ established in Fig.~\ref{fig4}, together with the weak dependence of mean strength on $\mathcal{S}$, allows these results to be extended to other levels of variability. In particular, changing $\mathcal{S}$ is approximately equivalent to rescaling the penalty parameter according to $\beta_{\rm equiv}=\beta\mathcal{S}/\mathcal{S}_0$, where $\mathcal{S}_0=0.06$ wt.\% is the reference scatter used in Fig.~\ref{fig6}. Thus, the optimal aging treatments identified here can be approximately mapped to other levels of composition variability without repeating the optimization.

The non-monotonic structure of the strength–variability trajectories shown in Fig.\ref{fig5} and \ref{fig6} might at first appear surprising, but we believe they can be understood in terms of the different time dependencies of mean strength and composition-induced variability already evident in Fig.\ref{fig3}. In particular, at 160 and 180$^\circ$C, the evolution of $W_{90}$ exhibits pronounced local maxima and minima that do not coincide with those of the mean strength. As a result, the trajectory through the mean strength–$W_{90}$ plane can change direction as aging progresses, producing distinct aging conditions with competing strength–variability characteristics. Near peak aging, where the mean strength varies slowly with time, continued changes in $W_{90}$ can produce substantial changes in the $\bar{\sigma}_y-{W}_{90}$ relationship. The abrupt transitions in optimal aging time observed in Fig.\ref{fig6} therefore arise from the selection between different regions of these underlying aging trajectories, rather than from discontinuities introduced by the optimization procedure. Physically, the different time dependencies of mean strength and variability reflect the combined effects of composition-dependent precipitation kinetics and the nonlinear relationship between precipitate populations and strengthening.

The results presented here demonstrate that robustness to composition variability is not determined solely by alloy chemistry, but can also be influenced through processing. In precipitation-hardened Al--Mg--Si alloys, heat treatments that maximize mean strength do not necessarily produce the most consistent mechanical response. Depending on aging temperature, improved robustness may be achieved through modest overaging or by selecting an underaged condition. The latter introduces an additional consideration, since continued precipitation during subsequent thermal exposure may alter the properties in service. Nevertheless, these results establish that heat treatment provides a means of accommodating composition variability, complementary to tighter control of feedstock chemistry.  Moreover, tools such as the one illustrated here could play an important role in allowing manufacturers to select optimal treatment schedules for different batches depending on composition so as to minimize the strength variation of the product. While not done today, this is something that could open doors for future expanded value of product with wider compositional variability.

An important consideration is that the appropriate measure of robustness depends on the intended application. Here, we have used the width of the strength distribution as a measure of property variability, but other measures may be more relevant when specific performance requirements must be satisfied. For example, 6xxx-series aluminium alloys are widely used in automotive energy-absorbing structures, where sufficient strength is required for energy absorption, but must be balanced against requirements for ductility and bendability \cite{Fourmann2024,Wang2026}. In such cases, both the lower and upper tails of the strength distribution may be important, and optimization could instead seek to maximize the fraction of material satisfying a prescribed strength window. Similarly, the weighted objective introduced here provides only a generic representation of the trade-off between mean strength and variability. More application-specific objectives could incorporate the economic value of the resulting product, including the dependence of its value on strength, property consistency, or the fraction of material meeting a given specification.

The approach also offers several opportunities for extension. The present analysis considers isothermal aging and independent variations in Mg and Si content in an Al--Mg--Si alloy. Extending the framework to non-isothermal heat treatments would allow more realistic processing schedules to be considered, but would require surrogate models trained to represent the effects of arbitrary thermal histories. Similarly, application to recycled alloys containing variations in additional elements, such as Cu or Zn, would require models capable of capturing their influence on microstructural evolution and properties. More broadly, the framework is not restricted to precipitation strengthening. Analogous approaches could be developed for other composition-sensitive properties, including corrosion resistance, fracture behaviour and surface quality, provided sufficiently well-calibrated, physically based models are available to support surrogate construction. In this sense, the central contribution is a route toward treating feedstock composition variability explicitly within materials and process design, rather than solely as a constraint to be minimized.

As the materials community moves toward increasingly variable feedstocks, maximizing nominal performance alone may no longer be sufficient; robust alloy and process design offers a framework for maintaining reliable mechanical performance in the presence of composition uncertainty.

\section{Acknowledgements}
CWS gratefully acknowledges financial support from the Natural Sciences and Engineering Research Council of Canada (NSERC) via the NSERC CREATE Net-Zero for Materials and Manufacturing Graduate Training Program, as well as support through a Visiting Professorship at Université Grenoble Alpes, hosted by the SIMaP laboratory.

\bibliography{cas-refs} 

\end{document}


\maketitle

\section*{Contents}

\begin{itemize}
    \item S1. Details of PreciSo and Strengthening Models
    \item S2. Gaussian Process Surrogate Construction

\end{itemize}

\newpage

\section{Precipitation and strengthening models}
\label{sec:preciso_strengthening}

\subsection{Precipitation model}

Precipitation kinetics were calculated using PreciSo, a class-based Kampmann--Wagner numerical (KWN) model [14]. The strengthening $\beta''$ phase was represented by spherical Mg$_5$Si$_6$ precipitates. Although $\beta''$ precipitates are needle-shaped, the spherical approximation was adopted to provide a simplified description of precipitation kinetics over the composition range considered here. Details of the governing equations and numerical implementation are provided by Perez \emph{et al.} [14].

Only Mg and Si were explicitly considered in the precipitation calculations. Contributions from other alloying elements and microstructural features were represented through a composition-independent baseline strength, $\sigma_0$.

The model requires the Mg and Si diffusivities, precipitate interfacial energy, solubility product, and atomic volumes of the matrix and precipitate. The diffusivities were expressed as $D_i(T) = D_{0,i} \exp\left(-\frac{Q_i}{RT}\right)$ while the solubility product was represented by

\begin{equation}
\log_{10}K=-\frac{A}{T}+B,
\end{equation}

where $A$ and $B$ were obtained from the Thermo-Calc TCAL4 database.

The material parameters used in the calculations are summarized in Table~\ref{tab:S1}.

\begin{table}[t]
\centering
\caption{
Material parameters used in the PreciSo precipitation and strengthening calculations. The interfacial energy and critical radius were determined through calibration against experimental aging curves.
}
\label{tab:S1}

\begin{tabular}{ll}
\toprule
Parameter & Value \\
\midrule

\multicolumn{2}{l}{\textit{Precipitation model}}\\
\midrule

Matrix atomic volume
    & $1.66\times10^{-29}$ m$^3$ \\

Precipitate atomic volume
    & $1.921\times10^{-29}$ m$^3$ \\

$D_{0,\mathrm{Mg}}$
    & $7.45\times10^{-5}$ m$^2$\,s$^{-1}$ \\

$Q_{\mathrm{Mg}}$
    & $120.5$ kJ\,mol$^{-1}$ \\

$D_{0,\mathrm{Si}}$
    & $6.90\times10^{-5}$ m$^2$\,s$^{-1}$ \\

$Q_{\mathrm{Si}}$
    & $117.6$ kJ\,mol$^{-1}$ \\

Solubility parameter, $A$
    & $16338.38$ K \\

Solubility parameter, $B$
    & $5.05534$ \\

\midrule
\multicolumn{2}{l}{\textit{Strengthening model}}\\
\midrule

$k_{\mathrm{Mg}}$
    & $342$ MPa \\

$k_{\mathrm{Si}}$
    & $137$ MPa \\

Shear modulus, $G$
    & $26$ GPa \\

Burgers vector, $b$
    & $0.286$ nm \\

Taylor factor, $M$
    & $3.06$ \\

\midrule
\multicolumn{2}{l}{\textit{Optimized parameters}}\\
\midrule

Interfacial energy, $\gamma$
    & $0.12147$ J\,m$^{-2}$ \\

Critical radius, $r_c$
    & $4.2071$ nm \\

Baseline strength, $\sigma_0$
    & $19.9$ MPa \\

\bottomrule
\end{tabular}
\end{table}

\subsection{Strengthening model}

The PreciSo calculations provide the evolution of matrix Mg and Si concentrations, precipitate size distributions, and precipitate volume fraction during aging. These quantities were used to calculate the solid-solution and precipitation contributions to yield strength.

Solid-solution strengthening was calculated using a Labusch-type relationship,

\begin{equation}
\sigma_{\mathrm{ss}} = M\left(k_{\mathrm{Mg}}c_{\mathrm{Mg}}^{2/3} + k_{\mathrm{Si}}c_{\mathrm{Si}}^{2/3} \right)
\label{eq:S_ss}
\end{equation}

where $M$ is the Taylor factor, $k_i$ is the strengthening coefficient for solute $i$, and $c_i$ is its concentration (in weight percent) remaining in the matrix [15]. The individual Mg and Si contributions were combined linearly.

Precipitation strengthening was calculated following [16], accounting for the full precipitate size distribution predicted by PreciSo. The precipitation contribution is

\begin{equation}
\sigma_{\mathrm{ppt}} = M\frac{Gb}{\bar{L}_s}\bar{\tau}^{*}
\label{eq:S_ppt}
\end{equation}

where $G$ is the shear modulus, $b$ is the Burgers vector, $\bar{L}_s$ is the effective precipitate spacing, and $\bar{\tau}^{*}$ is the mean normalized obstacle strength. The effective spacing was approximated by

\begin{equation}
\bar{L}_s = \bar{R} \sqrt{\frac{2\pi}{3f_p}}
\label{eq:S_spacing}
\end{equation}

where $f_p$ is the precipitate volume fraction and $\bar{R}$ is the number-weighted mean precipitate radius,

\begin{equation}
\bar{R} = \frac{\sum_i N_iR_i}{\sum_i N_i}
\end{equation}

Here, $R_i$ and $N_i$ are the radius and number density of precipitates in size class $i$.

To account for the precipitate size distribution, the three-dimensional distribution predicted by PreciSo was transformed into a distribution of precipitate intersection radii in the slip plane using a Wicksell-type stereological transformation [16]. The unnormalized distribution was calculated as

\begin{equation}
\widetilde{p}(r) = \frac{r}{\bar{R}}\sum_{R_i>r} \frac{N_i}{\sqrt{R_i^2-r^2}}
\label{eq:S_wicksell}
\end{equation}

where $r$ denotes the radius of the precipitate intersection with the slip plane. The normalized distribution is

\begin{equation}
p(r) = \frac{\widetilde{p}(r)} {\displaystyle \int_0^{R_{\max}}\widetilde{p}(s)\,ds}.
\end{equation}

The mean normalized obstacle strength was then obtained by integrating over this distribution,

\begin{equation}
\bar{\tau}^{*} = \int_0^{R_{\max}} \tau^{*}(r)p(r)\,dr
\label{eq:S_mean_tau}
\end{equation}

The normalized obstacle strength was evaluated as

\begin{equation}
\tau^{*}(r) =
\begin{cases}
0.9 \left(\dfrac{r}{r_c}\right)^{3/2}\left[1-\dfrac{1}{6}\left(\dfrac{r}{r_c}\right)^5\right],
& r<r_c \\[8pt]
0.75, & r\geq r_c
\end{cases}
\label{eq:S_tau}
\end{equation}

where $r_c$ is the critical radius defining the transition between shearable and non-shearable precipitate obstacles.

The solid-solution and precipitation strengthening contributions were combined linearly,

\begin{equation}
\sigma_y = \sigma_0 + \sigma_{\mathrm{ss}} + \sigma_{\mathrm{ppt}}
\label{eq:S_total_strength}
\end{equation}

where $\sigma_0$ represents contributions to yield strength not explicitly included in the precipitation and solid-solution strengthening calculations.

\subsection{Optimization of model parameters}

The precipitate interfacial energy ($\gamma$), critical radius ($r_c$), and baseline strength ($\sigma_0$) were calibrated against experimental yield-strength measurements reported by Bao [17]. These experiments were performed on experimental extrusions that were subsequently cold rolled then solution treated at 550$\circ$C for 5 minutes, quenched in water and immediately isothermally aged.  A single parameter set was determined by simultaneously fitting aging curves at 160, 180 and 200$^\circ$C. For each trial pair $(r_c,\gamma)$, PreciSo calculations were performed at the three aging temperatures. The corresponding baseline strength $\sigma_0$ was determined by minimizing a robust Cauchy loss function,

\begin{equation}
L = \sum_T\frac{1}{N_T}\sum_{i=1}^{N_T}c^2\ln\left[1+\left(\frac{\sigma_i^{\mathrm{model}}+\sigma_0-\sigma_i^{\mathrm{exp}}}{c}\right)^2\right]
\label{eq:S_optimization}
\end{equation}

where $N_T$ is the number of experimental measurements at temperature $T$, $\sigma_i^{\mathrm{model}}$ is the calculated strength excluding $\sigma_0$, $\sigma_i^{\mathrm{exp}}$ is the experimental yield strength, and $c=20$ MPa is the robust scale parameter. The normalization by $N_T$ gives equal weight to each aging temperature, while the Cauchy loss reduces the influence of large residuals. The global optimization over $(r_c,\gamma)$ was performed using differential evolution giving optimum values of $r_c = 4.2071$~nm, $\gamma = 0.12147$~J/m$^2$ and $\sigma_0$ = 19.9~MPa.

\begin{figure}[t]
\centering
\includegraphics[width=\columnwidth]{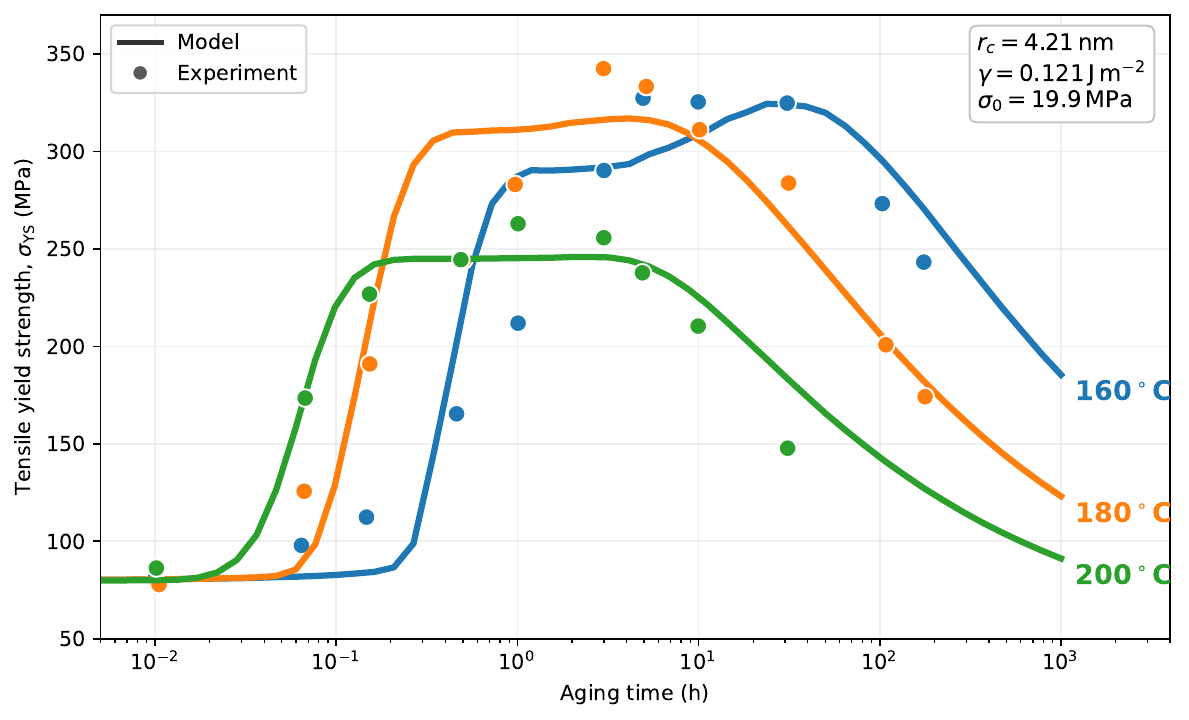}
\caption{Calibration of the PreciSo precipitation-strengthening model against experimental yield-strength measurements reported by Bao [17]. Symbols represent experimental measurements and solid lines represent model predictions during isothermal aging at 160, 180 and 200$^\circ$C. A single parameter set, $\gamma=0.12147$ J\,m$^{-2}$, $r_c=4.2071$ nm and $\sigma_0=19.9$ MPa, was obtained by simultaneous optimization against all three aging curves.}
\label{fig:S1}
\end{figure}

Figure~\ref{fig:S1} compares the corresponding model predictions with the experimental measurements. The model reproduces the principal features of the aging response, including the temperature dependence of strengthening kinetics, peak strength, and subsequent overaging. Some discrepancies remain in the timing and shape of the strengthening response, particularly during the transition from underaging to peak aging.

\section{Gaussian Process Surrogate Construction}

Separate Gaussian-process (GP) surrogate models were constructed for each aging temperature (160, 180 and 200$^\circ$C) to emulate the coupled PreciSo precipitation-strengthening model. Each surrogate predicts yield strength as a function of Mg content, Si content and aging time,

\begin{equation}
\sigma_{\rm YS}=f(c_{\rm Mg},c_{\rm Si},t),
\end{equation}

where $c_{\rm Mg}$ and $c_{\rm Si}$ are the Mg and Si concentrations (wt\%), respectively, and $t$ is the aging time (s). Since the aging simulations span several orders of magnitude in time, the GP was trained using $\log_{10}(t)$ rather than the raw aging time to improve numerical conditioning and interpolation accuracy.

\subsection{Initial training dataset}

The initial training dataset was generated using a Latin Hypercube Sampling (LHS) design over the prescribed Mg--Si composition domain. LHS provides an efficient space-filling sampling strategy that distributes training points uniformly throughout the design space while requiring relatively few simulations.

For each sampled composition, the PreciSo model was evaluated over the complete aging-time schedule used in this work. Each training sample therefore consisted of the three-dimensional input vector

\begin{equation}
\mathbf{x}
=
\left(
c_{\rm Mg},
c_{\rm Si},
\log_{10}t
\right),
\end{equation}

together with the corresponding PreciSo prediction of the yield strength,

\begin{equation}
y=\sigma_{\rm YS}.
\end{equation}

An initial LHS design consisting of 25 alloy compositions was used to construct the first surrogate model.

\subsection{Gaussian Process regression}

The surrogate was constructed using Gaussian Process regression with a Mat\'ern covariance kernel ($\nu=2.5$),

\begin{equation}
k(\mathbf{x},\mathbf{x}')
=
\sigma_f^2
k_{\rm Mat\acute{e}rn}
(\mathbf{x},\mathbf{x}'),
\end{equation}

where independent characteristic length scales were assigned to each input variable. The kernel hyperparameters, including the signal variance and characteristic length scales, were determined by maximizing the log-marginal likelihood of the training data.

In addition to predicting the expected yield strength, the GP provides an estimate of the predictive uncertainty at any point within the design space. This uncertainty estimate was subsequently used to guide adaptive refinement of the surrogate.

\subsection{Active learning}

Following construction of the initial surrogate, the GP was evaluated over a dense candidate grid spanning the composition--time domain. At each active learning iteration, the candidate point exhibiting the largest predictive standard deviation,

\begin{equation}
\mathbf{x}_{\rm next}
=
\underset{\mathbf{x}}{\arg\max}
\;
\sigma_{\rm GP}(\mathbf{x}),
\end{equation}

was identified and a new PreciSo simulation was performed at this location. A new PreciSo aging simulation was then performed for the corresponding Mg and Si composition, generating a complete aging curve.All points from this curve were added to the training dataset before retraining the GP.

Twenty active learning iterations were performed, resulting in a final training set comprising 45 unique alloy compositions. Since each composition was evaluated over the complete aging-time schedule, the final surrogate was trained using approximately 2800 precipitation-strength simulations.

Adaptive sampling naturally concentrates additional simulations in regions where the response surface varies rapidly or where training data are sparse, while avoiding unnecessary sampling in regions where the response is already well described. Active learning was terminated once the maximum GP predictive uncertainty over the candidate grid fell below a prescribed tolerance of 5 MPa.




\subsection{Software implementation}

All surrogate model development was performed in \texttt{Python} using the \texttt{GaussianProcessRegressor} implementation provided by the \texttt{scikit-learn} machine learning library. Numerical computations were carried out using \texttt{NumPy} and \texttt{SciPy}, data manipulation using \texttt{Pandas}, and visualization using \texttt{Matplotlib}. Custom Python routines were developed for Latin Hypercube sampling, active learning, uncertainty-guided sample selection, Monte Carlo uncertainty propagation and surrogate validation.